\documentclass[10pt]{article}
\usepackage{amsmath,subfig,graphicx,cite,fancyhdr}
\usepackage[affil-it]{authblk}
\usepackage[margin=1in]{geometry}
\begin{document}

\title{Multi-scale approach for analyzing convective heat transfer flow in background-oriented Schlieren technique}

%% Group authors per affiliation:
\author[1,2]{Gannavarpu Rajshekhar\thanks{Corresponding author: gshekhar@iitk.ac.in\\ 
DOI: 10.1016/j.optlaseng.2018.07.002\\
\copyright 2018 This manuscript version is made available under the CC-BY-NC-ND 4.0 license\\
http://creativecommons.org/licenses/by-nc-nd/4.0/
} }
\affil[1]{Department of Electrical Engineering, Indian Institute of Technology Kanpur, Kanpur-208016, India }
\affil[2]{Center for Lasers and Photonics, Indian Institute of Technology Kanpur, Kanpur-208016, India}
%\ead{gshekhar@iitk.ac.in}
\author[3,4]{Dario Ambrosini}
\affil[3]{DIIIE, University of L' Aquila, P. le E. Pontieri 1, 67100 L' Aquila, Italy}
\affil[4]{ISASI-CNR, Institute of Applied Science and Intelligent Systems, Via Campi Flegrei 34, Pozzuoli (NA) 80078, Italy}

%% or include affiliations in footnotes:
\date{}
\maketitle

\begin{abstract}
The paper introduces a multi-scale processing method for quantitative study and visualization of convective heat transfer using diffractive optical element based background-oriented schlieren technique. 
The method relies on robust estimation of phase encoded in the fringe pattern using windowed Fourier transform and subsequent multi-scale characterization of the obtained phase using continuous wavelet transform.
As the phase is directly mapped to the refractive index fluctuations caused by the temperature gradients, the multi-scale inspection provides interesting insights about the underlying heat flow phenomenon. 
The performance of the proposed method is demonstrated for quantitative flow visualization.
\end{abstract}

\section{Introduction}
For flow visualization, optical methods constitute an important class of measurement techniques because of non-invasive operation, full-field measurement and capability to provide both qualitative and quantitative insights with good sensitivity \cite{merzkirch2012flow,ambrosini2006optical}.
Some of the prominent optical techniques for flow visualization are outlined in \cite{prenel2012flow}.
Among these, beam deflection approach based on background distortion \cite{settles2012schlieren}  is relatively popular for flow visualization because of relative ease of operation, digital data processing, cost-effective design and robustness against external disturbances.
The applicability of this technique is further expanded by the enormous advances made in the field of computer hardware and software.
This technique relies on mapping the refractive index variation to the distortions experienced by a background pattern. 
Background distortion techniques include background oriented schlieren (BOS) \cite{raffel2000applicability,richard2001principle} and its variants \cite{sourgen2012reconstruction}, and synthetic schlieren technique \cite{dalziel2000whole}.
Over the years, similar methods based on grid pattern distortions have also been proposed \cite{massig1999measurement,perciante2000visualization}.
Most of these techniques are improvements over the schlieren method originally proposed by Schardin \cite{schardin1942schlierenverfahren}.
Despite the historical development, the name background-oriented schlieren technique is the most
common, and the BOS acronym is the best known. 
Updated reviews of the current state of the art in this field are given in \cite{raffel2015background,settles2017review}.

Recently, diffractive optical element (DOE) based BOS technique, strictly resembling Schardin's schlieren \#2 technique, was proposed for flow visualization \cite{ambrosini2007heat}. 
The technique relies on projecting a fringe pattern generated by the DOE element through a phase object or test section, and recording the deformed fringe patterns caused by heating the test section. 
Subsequently, fringe analysis methods \cite{rajshekar2012p} can be applied for fringe pattern demodulation or phase extraction. 
In a recent work \cite{ambrosini2012role}, application of fringe analysis methods based on Fourier transform method and windowed Fourier transform (WFT) method was studied and quantitatively compared with particle image velocimetry (PIV) method \cite{westerweel1997fundamentals} for flow visualization, and WFT method was shown to have superior performance with respect to other methods.
Though DOE based schlieren technique offers the advantages of compactness, robustness and good measurement sensitivity, the ability to selectively investigate or visualize the heat flow features or structures on a spatial scale is lacking. 
Enabling such capability would impart great value addition to the schlieren technique and be of great significance for quantitative heat flow visualization.

The main aim of this paper is to enhance the capabilities of the DOE based schlieren technique for flow measurements using robust fringe processing and multi-scale visualization methods. 
The experimental setup for the DOE based schlieren technique and the applied fringe processing method is discussed in section \ref{section_setup}. 
Subsequently, theory of  multi-scale analysis for the proposed technique is outlined in section \ref{section_theory}.
The results are presented in section \ref{section_result}.
These are followed by discussions and conclusions.

\section{Experimental Setup \& Fringe Processing}
\label{section_setup}
\begin{figure}[t]
    \centering
    \includegraphics[width=0.5\textwidth]{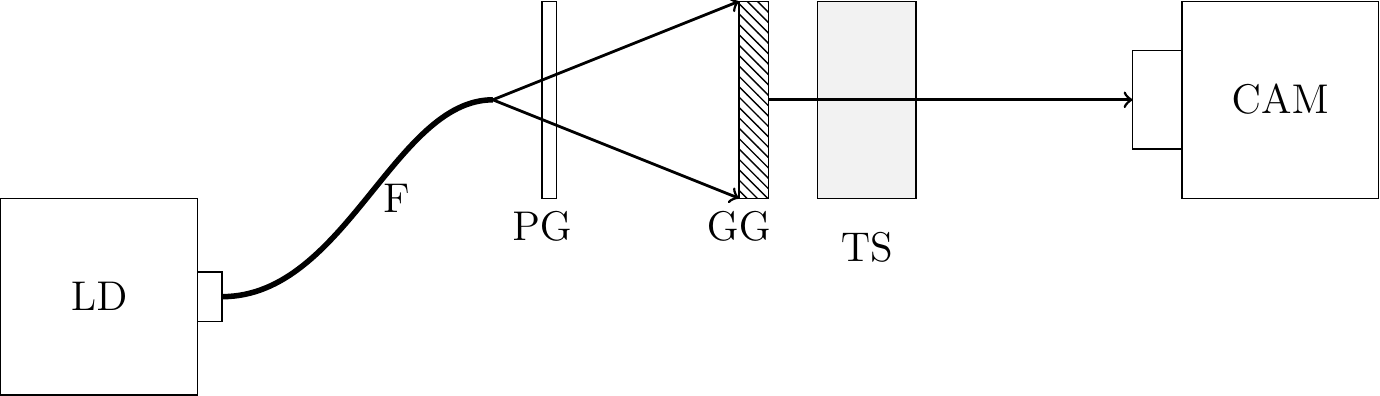}
    \caption{Schematic. LD: Laser Diode, F: Single mode optical fiber, PG: Phase grating, GG: Ground glass, TS: Test Section, CAM: Camera }
    \label{fig_setup}
\end{figure}

The schematic of the diffractive optical element based BOS setup is shown in Fig. \ref{fig_setup}.
The illumination source is a laser diode (Lasiris, wavelength $\lambda=638.5$ nm, output power 5 mW) which is pigtailed to a single mode fiber.
The fiber end, which acts as a point source, emits spherical waves which impinge on the diffractive optical element (DOE).
The DOE is a blazed phase grating realized on index-matching epoxy, and acts as a beam divider in our setup.
The beams emerging from the grating form interference fringes or fringe pattern in the superposition region.
The fringe period for the fringe pattern can be controlled by changing the distance between the fiber end and grating.
The fringe pattern is projected on a ground glass plate which acts as a screen.
The test section consisted of rib-roughened vertical channels (details described in \cite{ambrosini2005comparative}) and is placed in front of the ground glass plate.
The aluminum rib has a square section with a height of 0.00485 m.
The investigated phenomenon is the natural convection heat transfer with air as the convective fluid.
A TV camera, focused on the screen, is used to view the background pattern through test section. 
The opacity of the glass is not critical; as the role of the ground glass is to act as a screen for the fringes, traditional ground glasses can be used. 
More details about the setup are outlined in \cite{ambrosini2007heat}, while the effects of changing the period of the fringe pattern and the distance between the test section and ground glass are discussed in \cite{ambrosini2012role}.

In this setup, the beam passing through the test section is deflected due to the spatial variations or fluctuations in the refractive index induced by temperature gradient. 
The deflection angle is directly proportional to the temperature gradient \cite{ambrosini2007heat}.
As the fringes are seen through a medium (air) with a spatially varying index of refraction (because of the heat transfer phenomenon), the result is a displacement of the pattern, or, equivalently, a phase modulation.
Subsequently, the phase modulated or deformed fringe pattern recorded on the camera can be modeled as
\[ I(x,y)=a[1+\cos[2\pi f_x x + \phi(x,y) ]]  \]
where $a$ is the amplitude term, $f_x$ is the carrier frequency dependent on the grating period and $\phi(x,y)$ is the phase term introduced due to temperature gradients.

\begin{figure}[t]
    \centering
    \includegraphics[width=0.5\textwidth]{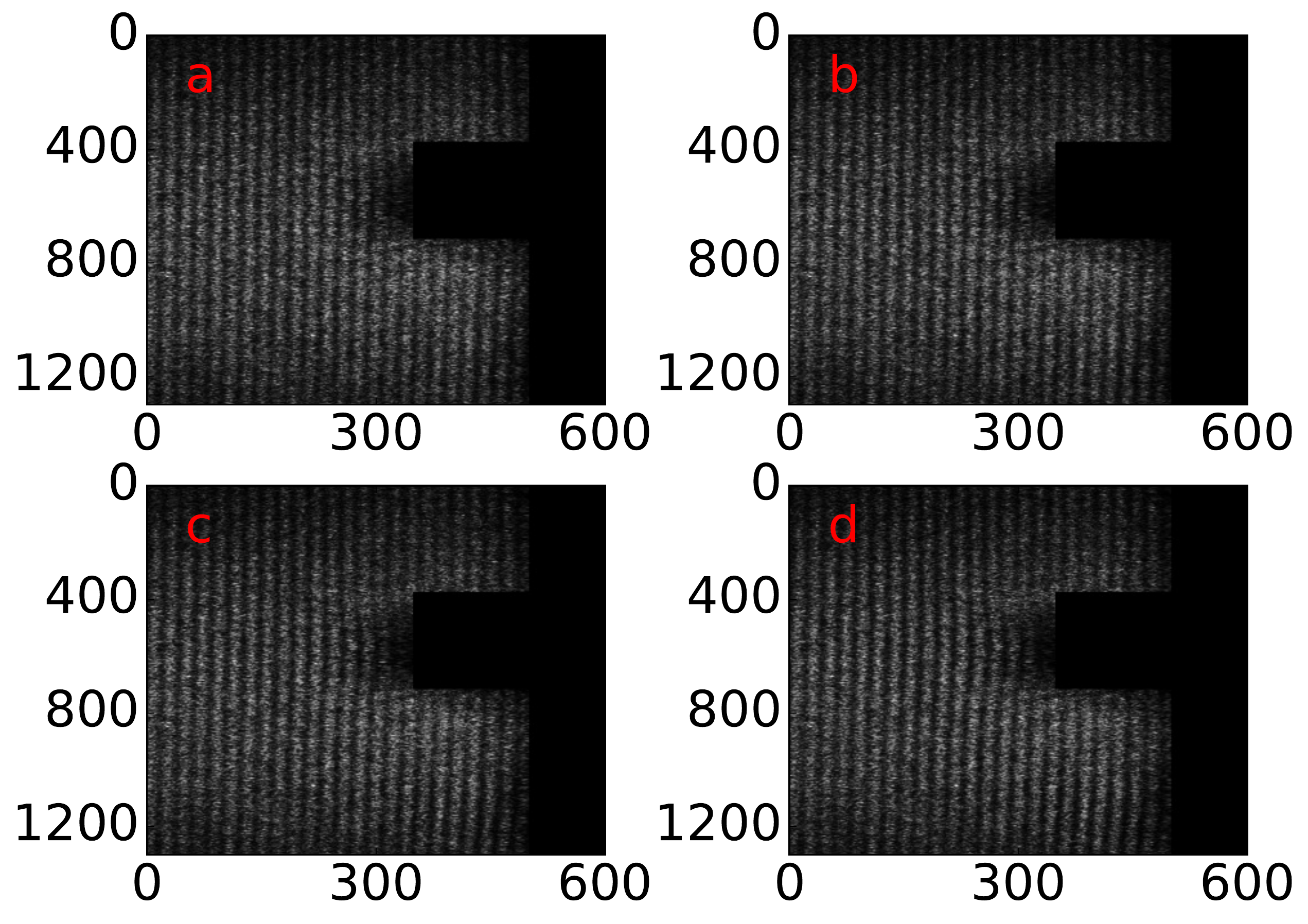}
   \caption{(a) Reference fringe pattern. Fringe patterns obtained for $\Delta T= 5^\circ C, 11^\circ C$, and $18^\circ C$ in (b-d). The values on the axes represent the number of pixels along the horizontal and vertical directions.}
   \label{fig_fringe}
\end{figure}

As a first step, we captured the fringe pattern corresponding to the initial state where no heating was applied to the rib. 
This reference fringe pattern is shown in Fig. \ref{fig_fringe}(a).
Note that we applied a mask of zeros for the region occupied by the rib in all images, as there are no fringes in this region. 
Subsequently, the rib was heated, and the refractive index changes caused by temperature gradient variations lead to a phase modulated fringe pattern.
Denoting the temperature variation with respect to the initial state as $\Delta T$, we recorded the fringe patterns corresponding to $\Delta T=5^\circ C$, $11^\circ C$ and $18^\circ C$, which are shown in Fig. \ref{fig_fringe}(b-d).

As the information about the temperature gradient is encoded in phase, the next step in our approach is fringe demodulation or reliable extraction of the embedded phase map from the deformed fringe pattern. 
For phase retrieval, we applied the windowed Fourier transform (WFT) method \cite{kemao2007two} which relies on local processing of the fringe pattern by using a two-dimensional block or window, which is usually selected to be a Gaussian function.
Effectively, the Fourier transform of the windowed signal is computed and analyzed, as opposed to the whole signal operation of the conventional Fourier transform.
The method enables local fringe processing and remedies the problems associated with the global nature of the Fourier transform.
More details about the implementation and working of the WFT method for processing DOE BOS fringes are presented in \cite{ambrosini2012role}.
For each $\Delta T$, the relative phase between the reference and the corresponding phase modulated fringe pattern was computed using the WFT method.

\section{Theory}
\label{section_theory}

The two-dimensional spatial phase distribution recovered in our setup is directly related to the temperature gradients, as mentioned before. 
The next task is to apply multi-scale analysis of the phase image to analyze the features at different scales or resolutions. 
This is performed using the two-dimensional continuous wavelet transform (2DCWT). For the phase function $\phi(x,y)$, the wavelet transform can be expressed as \cite{antoine1993image,torrence1998practical},
\begin{equation}
    W(x_1,y_1,\alpha)=\frac{1}{\alpha}\iint \phi(x,y)\psi^*{\left(\frac{x-x_1}{\alpha},\frac{y-y_1}{\alpha}\right)}\,dxdy
\end{equation}
where $\psi(x,y)$ is the mother wavelet, and $*$ denotes the complex conjugate. 
Also, $(x_1,y_1)$ indicates the translation of the mother wavelet in the 2D space and dilation is represented by the scale parameter $\alpha$. 
Effectively, the wavelet transform can be represented as the correlation of the signal $\phi$ with the translated and dilated version of the mother wavelet $\psi$. 
Regions where the wavelet transform values are large indicate high correlation between the signal and the wavelet.
Note that the angular dependence of the 2DCWT is not considered here for the sake of computational simplicity. 

Because of the correlation operation, the computation of the wavelet transform can be efficiently performed in the Fourier domain. The Fourier transform of the wavelet is given as
\begin{equation}
    \hat{W}(f_x,f_y,\alpha)=\alpha \hat{\phi}(f_x,f_y)\hat{\psi}^*(\alpha f_x,\alpha f_y)
\end{equation}
where $f_x$ and $f_y$ indicate the spatial frequencies, and $\hat{\phi}$ and $\hat{\psi}$ denote the 2D Fourier transforms of phase $\phi$ and mother wavelet $\psi$.
Hence, the 2DCWT can be computed using an inverse Fourier transform, 
\begin{align}
W(x_1,y_1,\alpha)=\displaystyle \iint& \alpha \hat{\phi}(f_x,f_y)\hat{\psi}^*(\alpha f_x,\alpha f_y)\\\notag
& e^{i2\pi(f_xx_1+f_yy_1)}\,df_xdf_y
\end{align}
By using fast Fourier transform algorithm for evaluating the Fourier transform, high computational performance can be achieved.

For our analysis, we chose the isotropic 2D Mexican hat function as the mother wavelet \cite{antoine1993image,antoine2008two},
\begin{equation}
        \psi(x,y)=(2-x^2-y^2)\exp\left[-{\left( \frac{x^2+y^2}{2}\right)}\right]
    \end{equation}

The main advantage of this wavelet is that it provides good spatial sensitivity for detecting sharp features in the image across all directions \cite{antoine1993image}. 
In addition, as the Mexican hat wavelet shown here is real and symmetric, the corresponding two-dimensional Fourier transform is also real and symmetric, which further simplifies the wavelet computations and also provides operational simplicity.

\begin{figure}[t]
    \centering
    \includegraphics[width=0.5\textwidth]{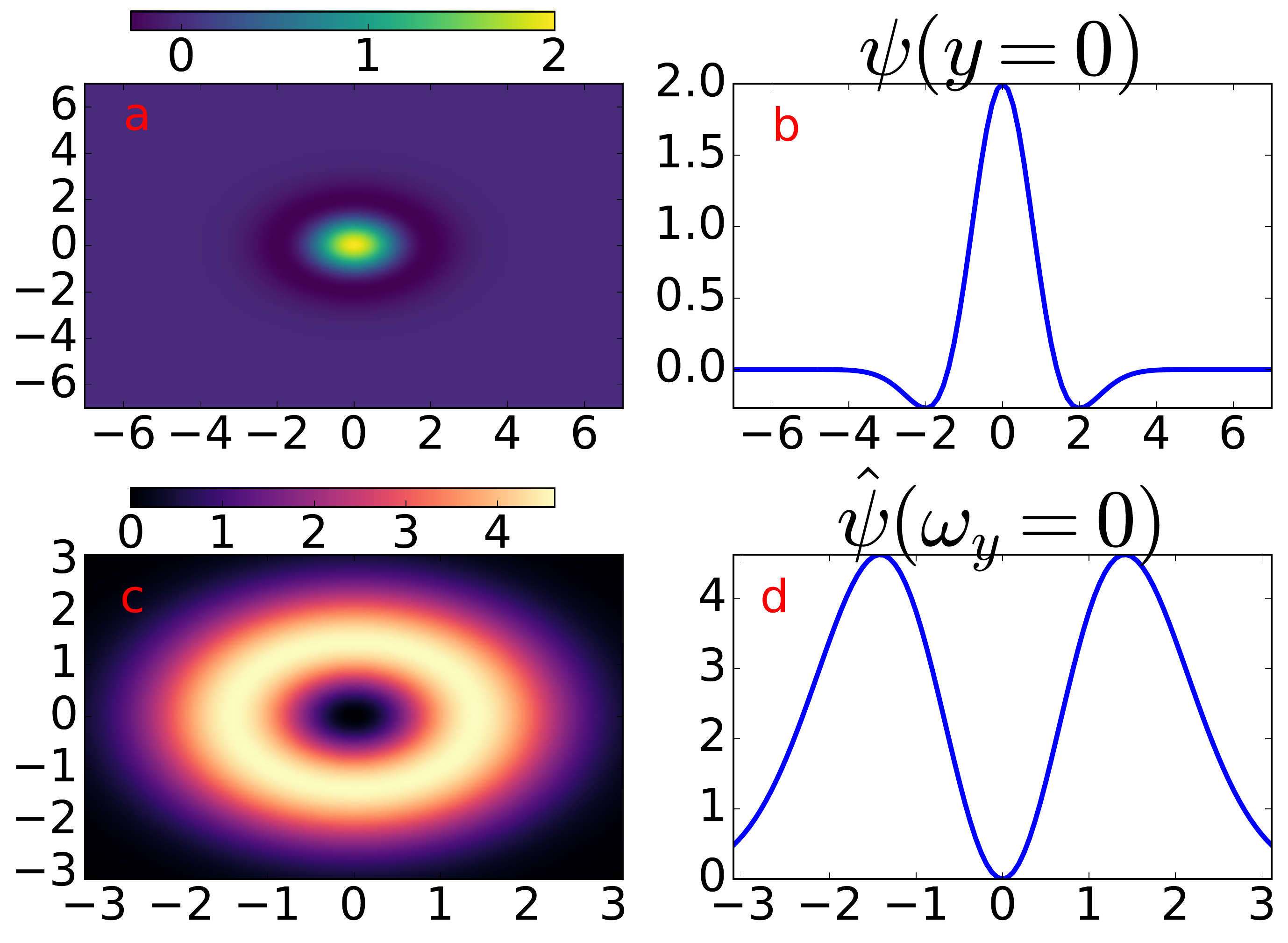}
    \caption{(a) 2D Mexican Hat mother wavelet. (b) Wavelet profile along $y=0$. (c) Fourier transform of the wavelet. (d) Fourier transform profile along $\omega_y=0$. }
    \label{fig_mexhat}
\end{figure}

The Mexican hat wavelet is shown in Fig.\ref{fig_mexhat}a. 
A line profile of the wavelet along $y=0$, i.e. $\psi(x,y=0)$ is shown in Fig.\ref{fig_mexhat}b. 
Similarly, the 2D Fourier transform of the wavelet $\hat{\psi}(\omega_x,\omega_y)$ is shown in Fig.\ref{fig_mexhat}c. 
Note that $\omega_x=2\pi f_x$ and $\omega_y=2\pi f_y$ are the angular spatial frequencies.
The line profile along $\omega_y=0$, or equivalently, $\hat{\psi}(\omega_x,\omega_y=0)$ is shown in Fig.\ref{fig_mexhat}d.

The scale parameter $\alpha$ has an inverse relationship with the spatial frequency [5].
For large values of the scale parameter, the dilated mother wavelet $\psi(x/\alpha,y/\alpha)$ is non-zero over a large range of $x$ and $y$, or equivalently exhibits large support.
As the wavelet transform is essentially a correlation operation, the dilated mother wavelet at large scales sees the phase signal as a whole. 
Due to this, sharper or finer details are not captured by 2DCWT for large $\alpha$.
On the other hand, for small values of $\alpha$, the support of the dilated mother wavelet is limited, and hence information about the finer details such as rapid phase variations or edges can be readily ascertained. 
Hence, by analyzing the wavelet transforms computed at different values of the scale parameter, the spatial features of the phase distribution, and consequently the temperature gradient can be monitored with varying levels of resolution.

\section{Results}
\label{section_result}
\begin{figure*}
    \centering
    \includegraphics[width=\textwidth]{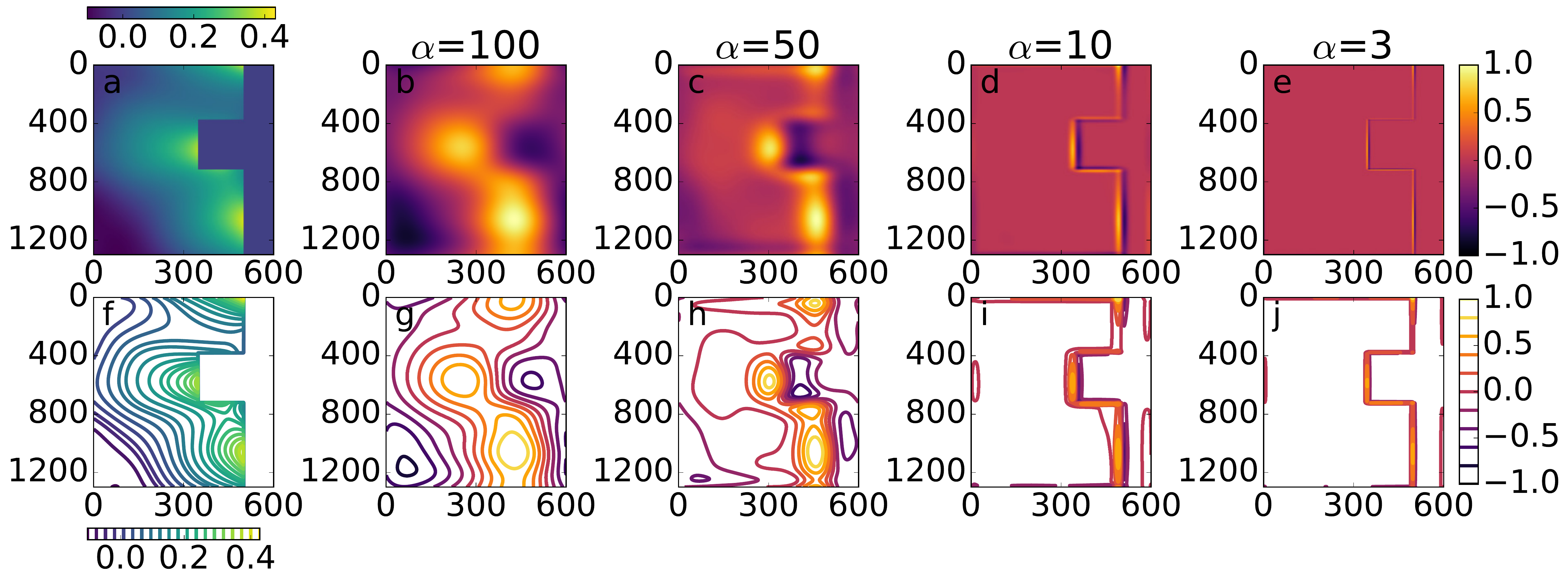}
    \caption{(a) Phase in radians for $\Delta T= 5^\circ C$ and its associated colorbar (in radians) on top. (b-e) Normalized continuous wavelet transform values of phase image for several values of scale and their respective common colorbar. (f) Contour plot of phase and its associated colorbar (in radians) beneath. (g-j) Contour plots of the wavelet transform values along with the respective common colorbar. 
 The values on the axes represent the number of pixels along the horizontal and vertical directions.}
    \label{fig_delta5}
\end{figure*}
\begin{figure*}
    \centering
    \includegraphics[width=1\textwidth]{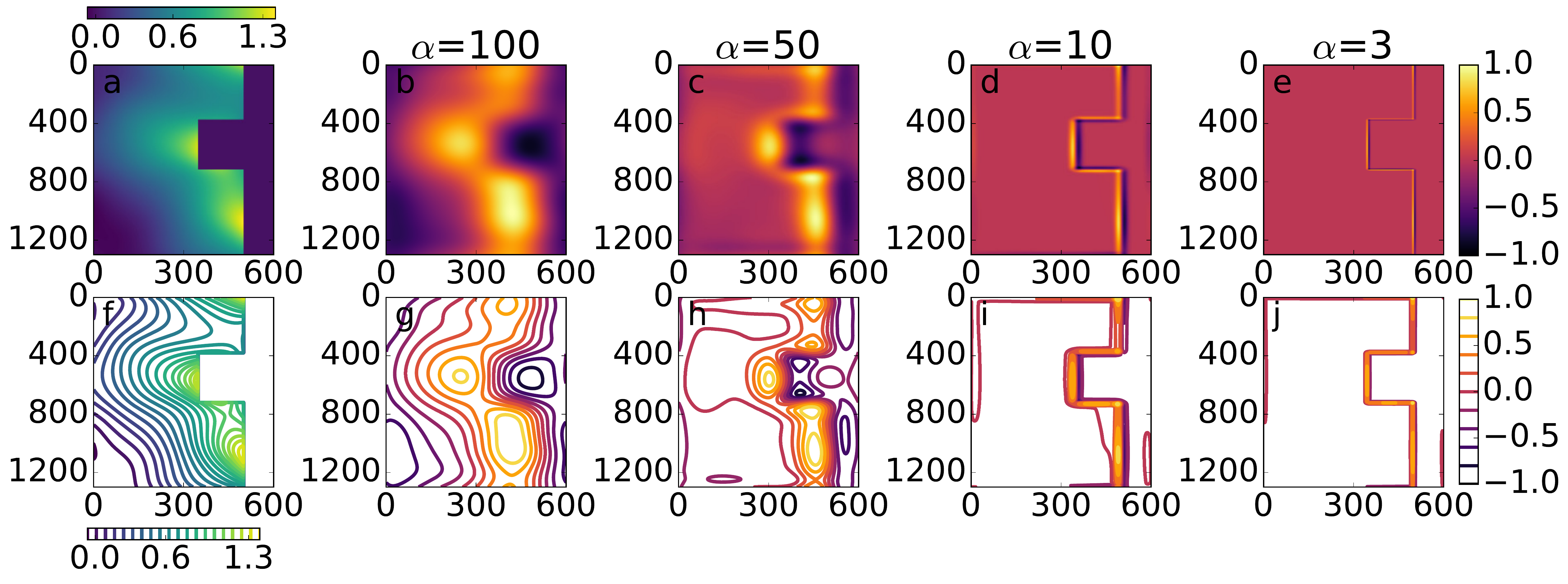}
    \caption{(a) Phase in radians for $\Delta T= 11^\circ C$ and its associated colorbar (in radians) on top. (b-e) Normalized continuous wavelet transform values of phase image for several values of scale and their respective common colorbar. (f) Contour plot of phase and its associated colorbar (in radians) beneath. (g-j) Contour plots of the wavelet transform values along with the respective common colorbar.
 The values on the axes represent the number of pixels along the horizontal and vertical directions.}
    \label{fig_delta11}
\end{figure*}

\begin{figure*}[h]
    \centering
    \includegraphics[width=1\textwidth]{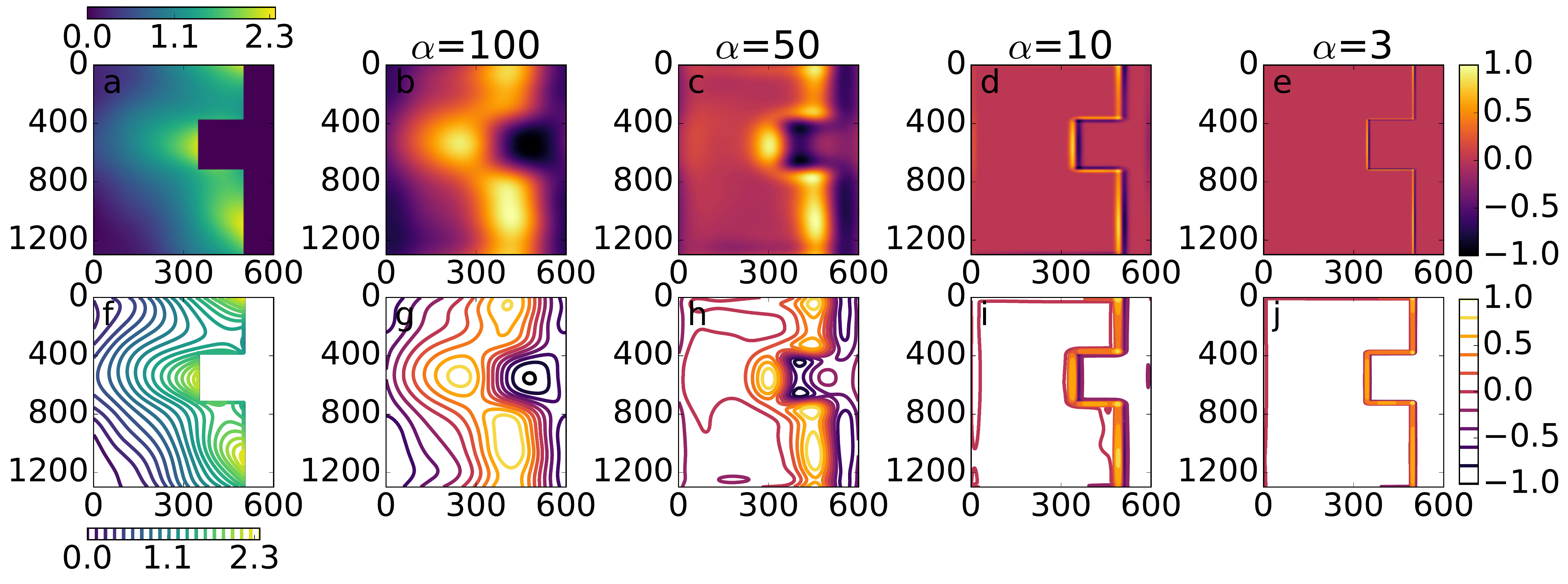}
    \caption{(a) Phase in radians for $\Delta T= 18^\circ C$ and its associated colorbar (in radians) on top. (b-e) Normalized continuous wavelet transform values of phase image for several values of scale and their respective common colorbar. (f) Contour plot of phase and its associated colorbar (in radians) beneath. (g-j) Contour plots of the wavelet transform values along with the respective common colorbar. 
 The values on the axes represent the number of pixels along the horizontal and vertical directions.}
    \label{fig_delta18}
\end{figure*}

For multi-scale analysis, we computed the 2DCWT at different scales in the range  $1\leq\alpha\leq 100$.
It needs to be noted that as the range of the wavelet transform values depends on the scale parameter, it is difficult to directly compare wavelet transforms computed at different scales.
Hence, we performed a normalization operation by dividing each wavelet image by its respective maximum value to limit the peak value to unity. 
This ensures uniformity in dynamic range across varying $\alpha$. 
Further, for ease of visualization, we applied a thresholding operation where the wavelet values lying within 1\% of the extremas  were ignored by substituting them with zeros.

For $\Delta T=5^\circ C$, the computed phase in radians using the WFT method and the corresponding contour plot are shown in Fig.\ref{fig_delta5}a and Fig.\ref{fig_delta5}f. 
In these plots, the variation in phase as we move away from the heating element or rib, is clearly evident, which signifies the temperature gradient change.
In the region close to the rib, the magnitude of phase is high, and it reduces gradually as the distance from the rib is increased.
For this phase image, the computed  2D wavelet transforms for decreasing values of scale parameter are shown in Figs.\ref{fig_delta5}b-e.
The corresponding contour plots of the 2D wavelet transforms are also presented in Figs.\ref{fig_delta5}g-j for the sake of clarity.

For high scale value of $\alpha=100$ in Fig.\ref{fig_delta5}b, the 2DCWT broadly indicates the presence of different structures or regions in the phase image. 
Especially, the region with high wavelet transform values indicate the presence of slowly varying (low frequency) but significant phase values (or equivalently temperature gradients). 
However, sharp structures such as the rib are not clearly visible, though the low 2DCWT values on the right side of the wavelet image indicate the presence of a different structure, which is essentially the masked region of the rib.   

Subsequently, as the scale parameter is lowered, increasingly finer details pertaining to the region near the boundary of the rib emerge in the 2DCWT images of Fig.\ref{fig_delta5}c-e. 
For scale values of $\alpha=50$ and $\alpha=10$, we can clearly see that most of the significant wavelet values are concentrated near the rib region.  
Thus, with these decreasing scales, we can selectively visualize the pockets of increasingly high frequency information, mainly occurring near the rib.
Finally, at a low scale value of $\alpha=3$, the sharp edge of the rib is clearly evident.
Thus, 2DCWT evaluated at low scales, has viable applicability for tracking the location of the rib.

We also repeated similar analysis for fringe patterns obtained at other temperatures. 
In Fig.\ref{fig_delta11}a and \ref{fig_delta11}f, we show the WFT computed phase image for $\Delta T = 11^\circ C$, and its contour plot. 
Note that with increasing temperatures, the range of phase values also increases which is evident from the colorbars (in radians) shown in Figures \ref{fig_delta5}(a), \ref{fig_delta11}(a) and \ref{fig_delta18}(a).
Importantly, the peak phase value, which occurs near the heated rib, increases as temperature is raised.
Similarly, the 2DCWT images at different scales are shown in Fig.\ref{fig_delta11}b-e, and their respective contour plots are shown in Fig.\ref{fig_delta11}g-j. 
Further, for $\Delta T=18^\circ C$, the phase image and its contour plot are shown in Fig.\ref{fig_delta18}a and \ref{fig_delta18}f. 
Similarly, the computed wavelet transform values are shown in Fig.\ref{fig_delta18}b-e, and the corresponding contour plots are presented in Fig.\ref{fig_delta18}g-j.
These figures also highlight the selective visualization of different structures and features of the phase image provided by the 2D continuous wavelet transform.
Consequently, these results clearly demonstrate the potential of multi-scale analysis for localizing and tracking structures in convective heat flow.

\section{Discussion}
In the paper, the proposed multi-scale analysis method allows us to selectively identify the different regions of interest from a single phase image with good sensitivity. 
The ability to analyze flow related structural information at different scales or resolutions could provide interesting insights for flow analysis. 
The multi-scale processing capability offered by the wavelet transform offers structure or feature detection from phase images.  
At large scale values or equivalently at low frequencies, the dominant wavelet spectrum values, as shown in Figs. \ref{fig_delta5}(b), \ref{fig_delta11}(b) and \ref{fig_delta18}(b), are located in regions corresponding to the dominant structures (with slow spatial variations) near the rib but the rib itself is not clearly identified. 
Conversely, at low scale values, the  wavelet spectrum clearly demarcates the rib region as shown in Figs. \ref{fig_delta5}(e), \ref{fig_delta11}(e) and \ref{fig_delta18}(e). 
This kind of selective visualization is the unique strength of wavelet based multi-scale analysis and could potentially be utilized for identifying structures such as plumes, thermals and heat source such as the rib. Regarding the limit of visualization, computing wavelet transform at very low scale values or equivalently at very high spatial frequencies might lead to aliasing errors related to sampling constraints. 

\begin{figure}[t]
    \centering
    \includegraphics[width=0.3\textwidth]{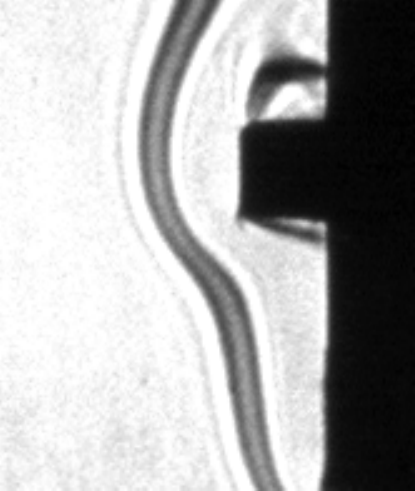}
    \caption{Typical focal filament schlieren image (details in \cite{ambrosini2005comparative})}
    \label{fig_7}
\end{figure}
Further insights can be obtained by comparing the present results with traditional focal filament schlieren technique \cite{ambrosini2005comparative,tanda1997natural}. 
Figure \ref{fig_7} shows a typical iso-deflection line (proportional to a iso-temperature gradient line) obtained by focal filament method on the same test section. 
By collecting many images of this type, a complete description of the temperature gradient field can be obtained, though at a great cost in experimental effort and time without the multi-scale capability.

By combining the techniques of windowed Fourier transform for phase extraction and wavelet transform for multi-scale analysis, we expect to provide a highly sensitive, robust and automated data processing method for flow visualization.
Overall, the advantages offered by our workflow are threefold: 
(1) sensitive and compact measurement methodology for quantitative flow measurements provided by the diffractive optical element based schlieren setup, 
(2) robust fringe processing and phase retrieval using windowed Fourier transform, and 
(3) multi-scale visualization, localization and feature detection of structures using continuous wavelet transform. 
\section{Conclusions}
The paper proposed a multi-scale data processing method in diffractive optical element based schlieren technique for visualizing convective heat transfer flow.
The continuous wavelet transform based approach is capable of identifying the structures and contours from phase maps obtained via fringe analysis in an automated fashion.
The proposed method fits nicely in the rapidly expanding field of advanced data processing methods for flow measurements.
The inclusion of the proposed method in the DOE BOS technique has the potential to significantly enhance the technique's applicability and utility for performing sensitive and quantitative flow measurements.

\section*{Acknowledgments}
Gannavarpu Rajshekhar gratefully acknowledges the funding obtained from Science and Engineering Research Board (SERB) India under grant ECR/2015/000180.

\end{document}